\documentclass{emulateapj}

\newcommand{\mpc}{\,\text{Mpc}\, h^{-1} }
\newcommand{\rhoth}{\rho_{\text{thres, HI}} /\langle \rho_{\rm H} \rangle}

\def\etal   {{et~al.}\ }

\slugcomment{To be submitted to ApJ}
\begin{document}

\title{The Topology of Reionization}
\author{Khee-Gan Lee, Renyue Cen, J. Richard Gott III and Hy Trac}
\affil{Department of Astrophysical Sciences, Princeton University, Princeton, New Jersey 08544, USA}
\email{lee@astro.princeton.edu}
\email{cen@astro.princeton.edu}
\email{jrg@astro.princeton.edu}

\begin{abstract}

Using the largest cosmological reionization simulation to-date ($\sim$24 billion particles) ,
we use the genus curve to quantify
the topology of the neutral hydrogen distribution on scales $\ge 1 \mpc$ as it evolves during cosmological reionization.
We find that the reionization process proceeds primarily 
in an inside-out fashion, where higher density regions become 
ionized earlier than lower density regions.
There are four distinct topological phases:
(1) Pre-reionization at $z \gtrsim 15$, when the genus curve is consistent with a Gaussian
density distribution.
(2) Pre-overlap at $10 \lesssim z \lesssim 15$,
during which the number of HII bubbles increases gradually with time,
until percolation of HII bubbles starts to take effect, 
characterized by a very flat genus curve at high volume fractions.
(3) Overlap at $8 \lesssim z \lesssim 10$, when large HII bubbles rapidly merge,
manifested by a precipitous drop in the amplitude of the genus curve.
(4) Post-overlap at $6 \lesssim z \lesssim 8$,
when HII bubbles have mostly overlapped
and the genus curve is consistent with a diminishing number
of isolated neutral islands.
After the end of reionization ($z \lesssim 6$), the genus of neutral hydrogen is consistent with Gaussian random phase, in agreement with observations. 

\end{abstract}

\keywords{cosmology: theory --- intergalactic medium --- large-scale structure of universe --- methods: numerical ---methods: statistical --- radiative transfer}

\section{Introduction}

Two recent major observational milestones
combine to suggest that the so-called ``dark ages" of the universe \citep{peacock92, rees98} 
may be rather complex. 
The first major milestone was laid down by 
the recent absorption spectrum observations of high redshift 
quasars from the Sloan Digital Sky Survey (SDSS),
indicating that the final reionization episode 
came to completion at $z\sim 6$
\citep[see][]{becker01,fan02,barkana02b,cen02,lidz02}.

The Wilkinson Microwave Anisotropy Probe (WMAP) 
polarization observations,
probing the ionization state of intergalactic medium (IGM)
in the high redshift universe,
set the second major milestone.
The WMAP observations indicate
$\tau_e=0.09\pm 0.03$ \citep{spergel06},
suggesting that the process of reionization occured at 
 redshifts considerably greater than $z \sim 6$,
possibly $z\ge 15$,
although the current uncertainty in $\tau_e$ allows
a very wide range of reionization histories.

Although the complex reionization process implied by these observations
did not appear as a total surprise
\citep[e.g., ][]{gnedin00b,
barkana01,madau02,
wyithe03,
mackey03,
cen03,
venkatesan03,
somerville03,
chiu03,
venkatesan03b,
ricotti04},
it is unclear at this time 
how the universe had actually evolved from $z\sim 1000$ to $z=6$,
for different scenarios could be designed to arrive
at the same observed total Thomson optical depth \citep[e.g., ][]{holder03}
as well as satisfy the sudden rise in ionizing radiation 
background observed at $z\sim 6$.

Thus, at present, many details of the cosmological reionization process are still unknown.
It is expected, however, that upcoming observations at various observational
bands from radio to optical promise to provide much more refined statistical
information than what is available today. Therefore, advances on the
theoretical side
are demanded in order to provide both useful insights
and a quantitative framework for the interpretation of future observations.


At present, we make the first attempt to quantify
the topological evolution of cosmological reionization,
in the context of the conventional scenario
that stellar radiation is primarily responsible for reionizing the universe.
We utilize a state-of-the-art cosmological 
reionization simulation of a box size $100 \mpc$, 
which involves a hybrid algorithm inclusive of detailed radiative transfer \citep{trac06b,shin07}.
This paper is organized as follows: 
the method section describes the genus and simulation code, 
followed by a description of the results and the manner in which they are presented. 
We then give a detailed discussion of the changes in the topology 
of the neutral fraction at different stages of reionization.

\section{Method}
\subsection{Genus: A quantitative measure of topology}

In order to study the topology of a three-dimensional data set like those in the simulation, 
we must first specify two-dimensional contours. 
From a continuous density distribution, 
this can be done by defining contours of constant density
at different volume fractions $f$, i.e.\ the 
fraction of the volume occupying the high-density side of the contour.
The topology of contours at a given threshold density 
can then be quantified with the genus. 

For a Gaussian distribution, 
\begin{equation}
f = \frac{1}{\sqrt{2 \pi}}\int^{\infty}_{\nu} e^{-x^2 / 2} {\rm d}x 
\label{vfrac}
\end{equation}  \citep{hamilton86}, where $\nu = \delta/\xi(0)^{1/2}$
is the number of standard deviations by which the threshold density $\delta$ 
of the contour departs from the mean density in 
a Gaussian distribution with covariance function $\xi(r)$. 
For example, the density contour with $\nu=1.5$ 
is that which encloses $\sim 93\%$ of the volume, 
or $f=0.07$ (i.e.\ 7\% of the volume has a higher density than $\nu = 1.5$).

While the relation between $f$ and $\nu$ as defined in Equation~\ref{vfrac} 
holds only for a Gaussian distribution, it has become the
convention to use $\nu$ as a convenient label 
for the different density contours
when studying the genus of a distribution, 
even if the distribution is not Gaussian.
In order to allow comparisons with the
published literature dealing with the genus,
we will mostly follow this convention
of referring to contours using $\nu$ as a label, 
which is related to the actual volume fraction $f$ by Equation~\ref{vfrac}.
This also has the advantage that large values 
of $\nu$ intuitively correspond to high threshold densities 
and vice-versa, whereas $f$ decreases with increasing threshold density.

Mathematically, the topology of an object is defined by its homeomorphism, 
i.e. its geometric properties under deformations and 
not transformations which `break' an object or 
`connect' it with others \citep[see, e.g.~][]{seifert:top}. 
One quantitative measure of an objects topology 
is its genus \citep{hamilton86,gott86,gott87}. 
We define the genus of a repeating or infinite contour as
\begin{equation}
G_s = (\mathrm{no.\;of\;holes}) - (\mathrm{no.\; of\; isolated\; regions}) \label{intgenus}
\end{equation} where a `hole' is like that in a torus, while an `isolated region' can be either above or below the threshold density 
of a contour \citep{gott86}. 
In this paper we define `holes' and `isolated regions' depending on whether the threshold density of the contour being studied is higher or lower than the 50\% median density value: for voids, isolated regions are defined as those with lower densities than the threshold density,
 while the reverse is true for dense pockets. 

Thus, a single dense region in an empty box and 
a single void in a solid box are both counted as 
`isolated regions' and are hence topologically equivalent. 
This allows for a symmetry in the definition of the 
genus for low density ($\nu < 0$) and high density ($\nu > 0$) regions. 

The genus of a repeating or infinite two-dimensional contour can be calculated 
by integrating its Gaussian curvature $K$ over the entire area $A$, 

\begin{equation}
G_s = \frac{-1}{4 \pi}\int K {\rm d}\!A \label{formgenus}
\end{equation} where the Gaussian curvature is defined as the reciprocal of the two principle radii of curvature $a_1$ and $a_2$  at the point, i.e. $K\equiv1/(a_1 a_2)$ \citep{hamilton86,gott86}. 
As an illustration, take for example a density contour which comprises of 20 isolated spheres, each of radius $r_s$. 
The Gaussian curvature at any point on the surface of each sphere is $K=r_s^{-2}$, while the total surface area is of course $20(4 \pi r_s^2)$. 
Using Equation \ref{formgenus} gives a genus of -20 as we would expect from the  definition given in Equation \ref{intgenus}. 
A multiply-connected, sponge-like contour would have a positive genus value, since the two principal radii of curvature on its surface tend to point in opposite directions rather than in the same direction, giving it a negative curvature. 
In a three-dimensional data set where each data pixel is essentially a cube, the genus is calculated by summing up the angle deficits at the vertices of the Lego-like contour surface. 
We use a computer program, {\sc Contour3D}, written by \citet{weinberg88} for this purpose. 

A point worth noting is that since the genus is calculated over the entire volume of the box, a combination of multiply-connected structures and isolated pockets could lower the amplitude of the genus. We thus have to be careful in distinguishing between, say, a volume with one figure-8 shaped object plus three small isolated spheres in it, as opposed to another volume with just one large isolated sphere, since both situations will give a total genus of -1.

The genus provides a useful tool for probing the relative connectedness of high- and low-density regions in the large-scale structure of the universe, for scalar quantities such as the dark matter distribution or density of HI. In addition to being dependent on the underlying distribution, the topology of large-scale structure also depends on the density threshold specified for a given contour surface. In general, contours of low threshold densities (i.e. voids occupying lowhigh volume fractions) have a different topology from contours of high threshold density (i.e. dense regions occupying low volume fractions) within a given data set. This provides an additional degree of freedom with which to study the topology of a density distribution, since we can now study the systematic variation of $G_s(\nu)$ with threshold densities using `genus curves'. 

In addition, the genus is an useful cosmological statistic for another reason. 
The standard big bang-inflationary model predicts that 
small-amplitude density fluctuations arise from a random phase 
Gaussian distribution \citep{bardeen83,bardeen86}. 
For such a distribution, the mean genus per unit volume $g_s \sim G_s / V$ is given by 
a simple analytic expression \citep{doroshkevich70,adler81,hamilton86}:

\begin{equation}
g_s = N(1-\nu^2)\,e^{-\nu^2/2} \label{gaussgenus}
\end{equation} where the normalization $N$ as function of wavenumber $k$ is given by 

\begin{equation}
N = \frac{1}{4 \pi^2}\left(\frac{\langle k^2 \rangle}{3} \right)^{3/2} = \frac{1}{4 \pi^2} \left(
\frac{ \int k^2 P'(k)\; {\rm d^3}\! k }{3 \int P'(k)\; {\rm d^3}\! k} \right)^{3/2} \label{genusamp}
\end{equation} where $P'(k) = P(k)\,e^{-r^2/2 r_{\rm sm}^2}$ is the power spectrum which has been smoothed with a smoothing radius $r_{\rm sm}$. 
We thus see that any Gaussian random-phase density distribution would have a curve of $g_s(\nu)$ with the same distinctive shape regardless of the power spectrum, which enters only through the amplitude. 
The implication is that the shape of the genus curve would remain unchanged through linear evolution of the fluctuations.
The initial conditions of our simulation at $z \sim 25$ display this characteristic curve (top left panel, Figure \ref{genuscurves}), and it is virtually indistinguishable from the theoretical curve . 

Equation \ref{gaussgenus} also shows that there is a symmetry between positive and negative $\nu$ in a random phase distribution, i.e. high- and low-density regions with the same absolute value of $\nu$ would have an identical topology. For $\vert \nu \vert > 1$ (i.e. $f < 16\%$ or $f > 84\%$) , the high- and low-density regions would comprise of isolated dense pockets and voids respectively as indicated by the negative genus. The median density contour with $\nu=0$ or $f=0.5$, on the other hand, has a positive genus indicating a sponge-like, multiply connected structure \citep{gott86,gott87}.

Historically, the genus has provided strong evidence in support of 
the Gaussian random phase hypothesis (and hence the now-standard inflationary big-bang model). 
Studies on the topology of large-scale structure as observed in galaxy surveys 
have yielded genus curves in remarkable agreement 
with the form shown in Equation \ref{gaussgenus}, 
apart from small-scale deviations due to non-linear gravitational effects and biasing 
from galaxy formation \citep{gott89,moore92,vogeley94,canavezes98,gott06}. 
In the present context, this means that the genus is a powerful and sensitive probe 
of the topology of the neutral fraction in the universe, 
as reionization causes a departure from a Gaussian random phase distribution.  

\subsection{Simulations}

We use the largest simulation of cosmic reionization run to date \citep{trac06b,shin07}.  It was run with a hybrid code containing a N-body
algorithm for dark matter, prescriptions for baryons and star formation, and
a radiative transfer (RT) algorithm for ionizing photons.  Here we summarize
the main simulation parameters.

The hybrid simulation was run with the cosmological parameters:  $\Omega_m=0.26$,
$\Omega_l=0.74$, $\Omega_b=0.044$, $h=0.72$, $\sigma_8=0.77$, and $n_s=0.95$, based on
the latest results from WMAP, SDSS, BAO, SN, and HST \citep[see][and
references therein]{spergel03}.  In a $100 \mpc$ simulation box, a
high resolution N-body calculation for $2880^3$ (i.e. 23.9 billion) dark matter particles on an
effective grid with $11520^3$ cells is performed using a particle-multi-mesh
code \citep{trac06a}.  With a particle mass resolution of
$3.02\times10^6\ \mathrm{M}_\odot h^{-1}$, halos can be reliably resolved down to masses
of $\sim10^8 \mathrm{M}_\odot h^{-1}$, 
accounting for the majority of photoionizing
sources.

The radiative transfer of ionizing radiation was run simultaneously with the
N-body calculations using a RT grid with $360^3$ cells.  However, the
ionization and recombination calculations were done for each particle
individually rather than on the grid to preserve small-scale information
down to scales of several comoving kpc~$h^{-1}$.  For post-processing, the dark
matter, baryons, and radiation are collected on a grid with $720^3$ cells
and the data is saved every 10 million years from $z=25$ down to $z=5$.

\section{Results}

\begin{figure}
\plotone{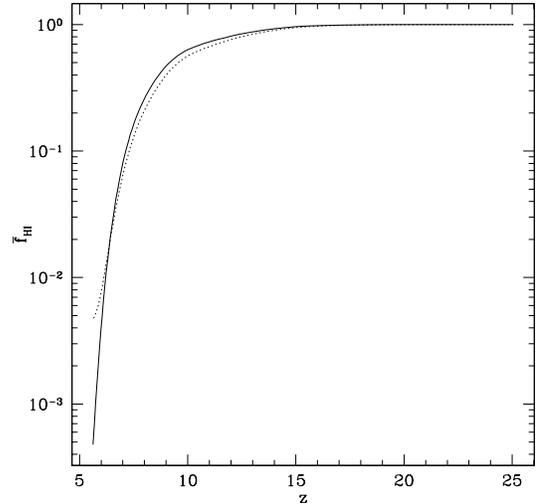}
\caption{ The volume-weighted (solid) and mass-weighted (dotted)  
 mean HI fractions from our simulation plotted as a function of redshift. 
These enable our results to be applied to other reionization
histories by matching the mean HI fractions at different epochs. 
  }
\label{fHIplot}
\end{figure}

In the present work, we study the density distribution of neutral hydrogen
(HI) at various redshifts during the epoch of reionization.   The
redshift evolution of the volume- and mass-weighted HI fractions are shown in
Figure~\ref{fHIplot}.  In our simulation, we can not resolve the damped
Lyman alpha systems associated with cooled neutral gas in halos above the
cooling mass.  Furthermore, we can only account for Lyman limit systems in
the form of neutral gas in partially self-shielded mini-halos down to about
half a dex in mass below the cooling mass.  Thus our HI densities are
representative of the IGM and partially exclusive of collapsed regions.  We
have chosen values for the star formation efficiency and radiation escape
fraction such that complete reionization is achieved by $z \sim 6$, as
suggested by observations of Lyman alpha absorption in high redshift
quasars.  The results of our analysis can also be applied to other
reionization histories since recent work \citep{Zahn07, McQuinn07} have
shown that the distribution of HI is, in general, similar for different
reionization histories when compared at the same ionization fraction.  We
thus use the volume-weighted ionization fraction $\bar{f}_{\rm HI}$ in our
plots as an alternative label for different epochs in addition to the
redshift $z$.

The $100 \mpc $ simulation box has $720^3$ cells, i.e. a spatial resolution of $\sim 0.14 \mpc$. 
Before beginning the analysis, we first smooth the data via a Fourier transform with a Gaussian kernel $W = e^{-r^2/2 r_{\rm sm}^2}$. 
We use a smoothing length of $r_{\rm sm}=1\mpc$, a scale which is adequate to smooth over individual pixels without washing out the structure of HII bubbles during reionization.

\begin{figure*}
\plotone{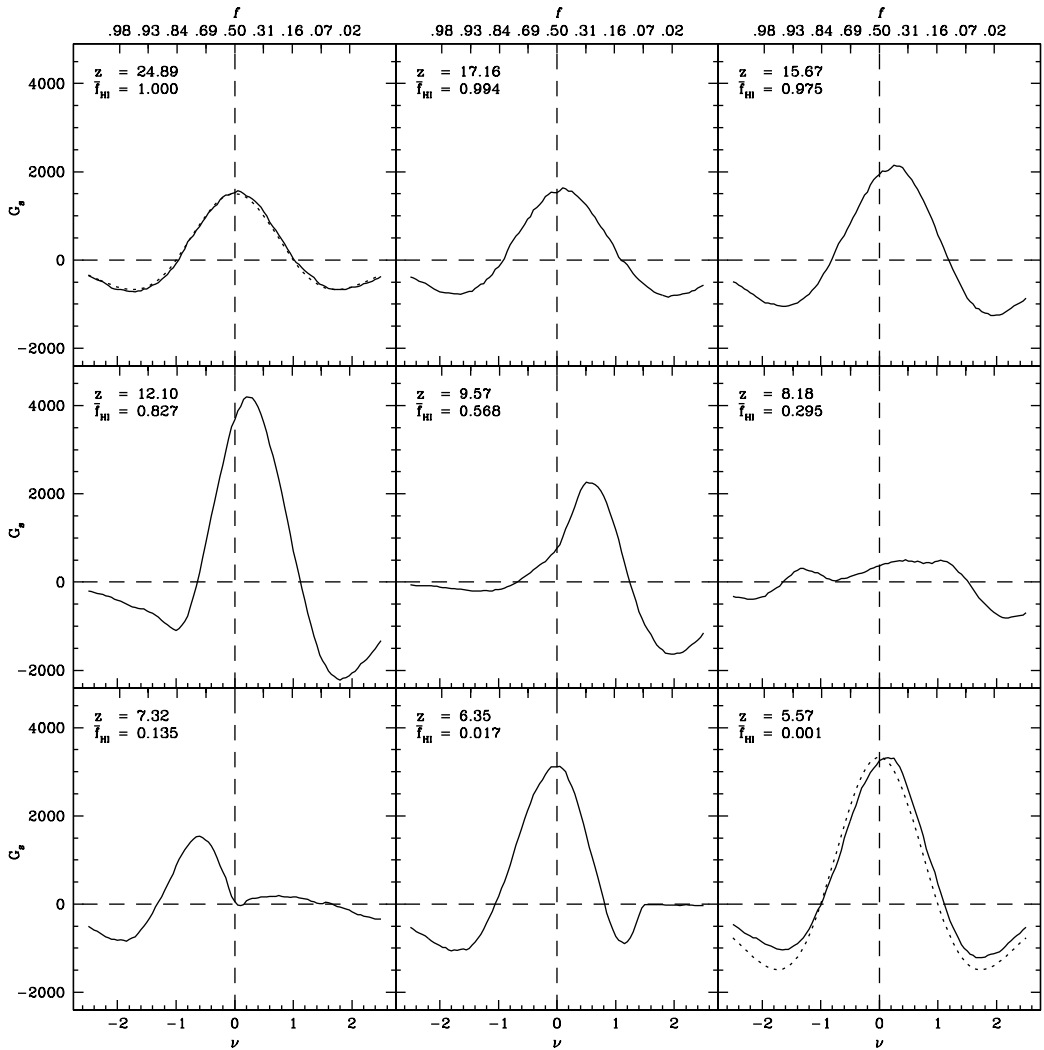}
\caption{Genus curves of the HI density distribution 
at various epochs during reionization, 
labeled by redshift $z$ and corresponding 
volume-weighted mean neutral hydrogen fraction $\bar{f}_{\rm HI}$. 
Following convention, we label the bottom axes with 
the quantity $\nu$ which is 
related to the volume fraction $f$ (on top axes) 
by Equation~\ref{vfrac} (see discussion in text).  
The dotted curves plotted in the top-left and bottom-right panels are 
the best-fit theoretical curves for a Gaussian random-phase distribution (Equation \ref{gaussgenus}). 
The vertical and horizontal dashed lines provide 
fiducial points for $\nu = 0$ and $G_s = 0$ respectively.
}
\label{genuscurves}
\end{figure*}

We apply the {\sc Contour3D} code to the simulation results at redshifts ranging from $z \sim 25$, well before the reionization process, to $z \sim 5.5$, after the end of reionization. 
Figure~\ref{genuscurves} shows the genus curves of the 
HI distribution at 9 different redshifts during reionization.
 These particular redshifts are shown as they represent the
`turning points' in the evolution of the topology, 
and the change between any two adjacent redshifts
plotted in Figure~\ref{genuscurves} is approximately monotonic.

Before we present the detailed discussion, 
we note that the actual HI threshold density $\rho_{\mathrm{thres, HI}}$ 
corresponding to a given volume fraction will not, 
in general, be constant as the universe evolves. 
As there is a spatial dependence 
on when neutral regions get reionized, 
an initially under-dense HI region could very well become a dense region 
relative to the rest of the reionized universe at a later time. 
We thus need to keep track of the redshift and spatial variation of 
the threshold density values in order to correctly interpret the changes in topology.

\begin{figure*}
\plottwo{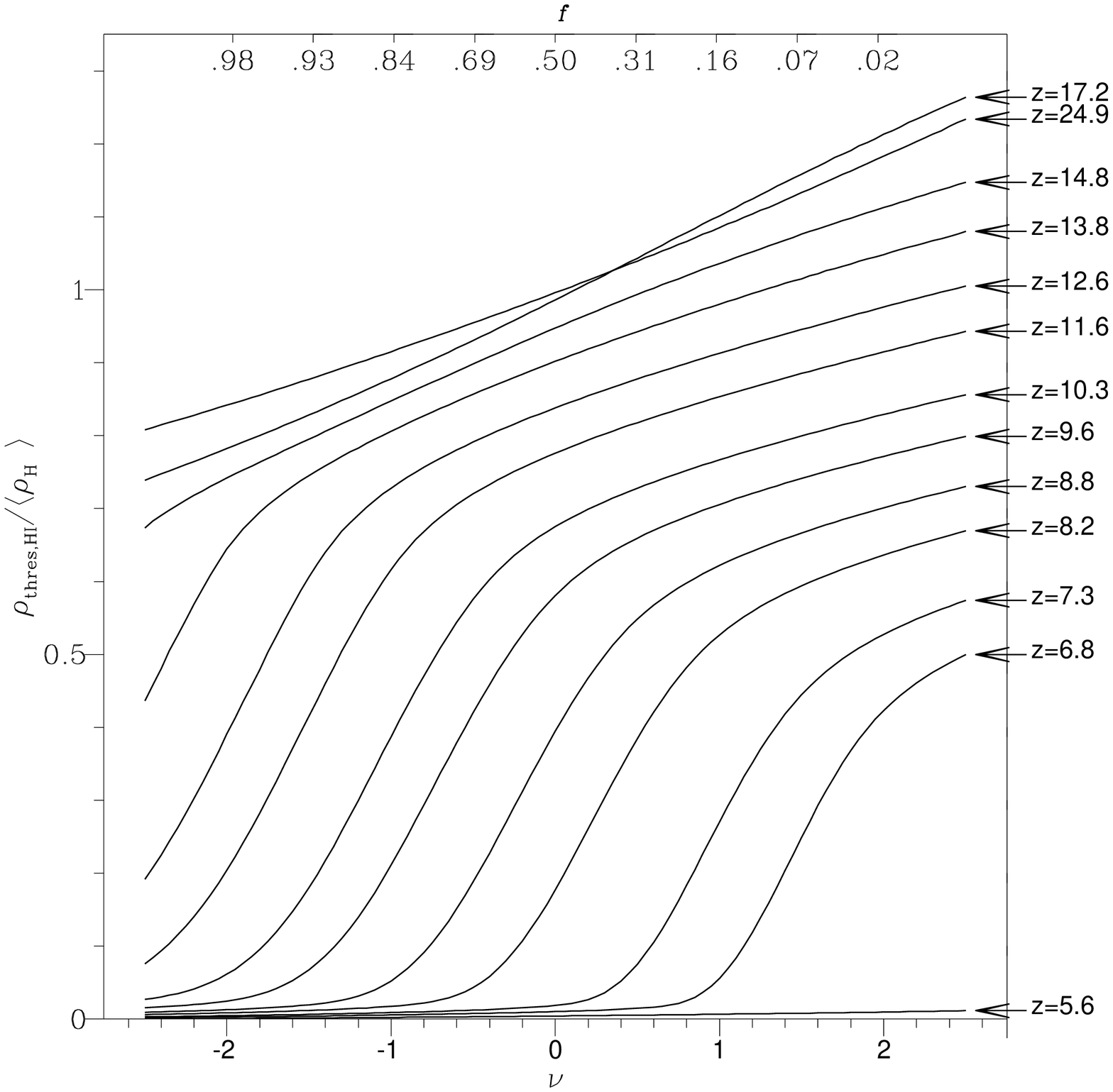}{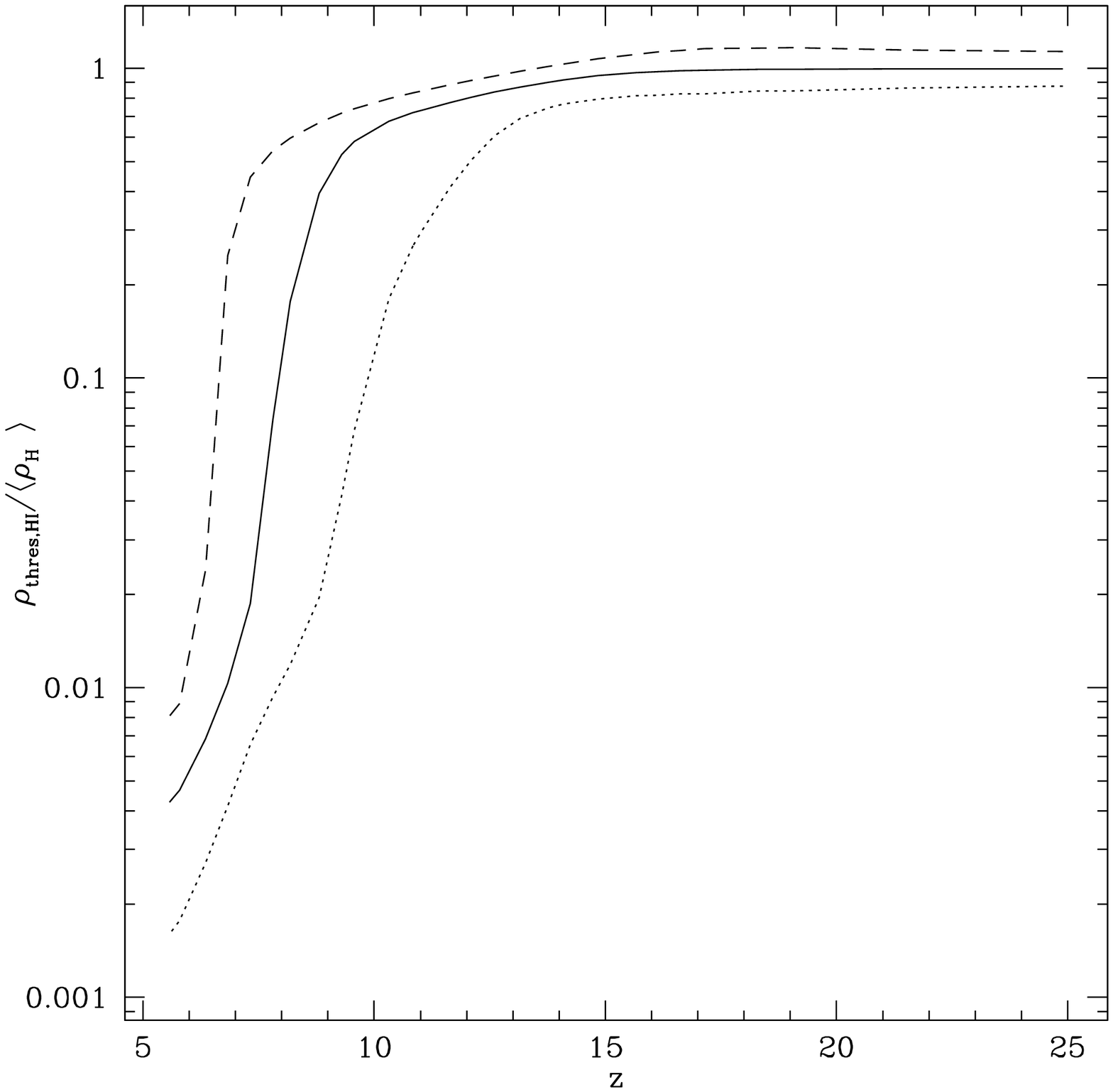}
\caption{(a)Left panel shows the contour threshold densities as a function of  $\nu$, (or equivalently volume fraction $f$; see Equation \ref{vfrac}), at different redshifts. Note that the threshold densities do not decrease monotonically in the early phases, between $z\sim 25$ and $z\sim 17$ (see discussion in text) (b) Right panel shows the redshift evolution of the threshold densities at $f=0.93$ or $\nu=-1.5$ (dotted line), $f=0.5$ or $\nu=0$ (solid line) and $f=0.07$ or $\nu=1.5$ (dashed line).  }
\label{rhoc}
\end{figure*}

For example, in the initial conditions at $z = 25$ before 
the onset of reionization, 
the median HI density at $\nu=0$ or $f=0.5$ is 
essentially the same as the overall mean hydrogen density , 
i.e. $\rhoth \approx 1$ 
 (recall that for a Gaussian distribution, 
$\nu$ is the number of standard deviations from the mean, 
so the median 50\% volume fraction threshold density 
is the same as the mean density at $\nu = 0$)  . 
By the end of reionization at $z\sim 6$, 
the average neutral fraction is close to zero, 
so $\rhoth \approx 0$. 
While we expect the densities in general to fall as reionization progresses, 
from Figure \ref{rhoc} it is obvious that $\rho_{\mathrm{thres, HI}}$ 
for different values of $\nu$ decrease at different rates during reionization, 
and at different redshifts. We thus need to keep track of the evolution 
and spatial variation of the actual threshold density values 
corresponding to a given $\nu$ in order to correctly interpret the changes in topology.

Another property we can study with the genus analysis is the characteristic 
scale of the neutral structures at a given epoch. 
From Equations \ref{gaussgenus} and \ref{genusamp}, 
the amplitude $N$ of a Gaussian distribution's genus curve 
 is related to the cube of its root-mean-squared (r.m.s.) wavenumber $\langle k^2 \rangle^{1/2}$. 
We thus have an estimate of the characteristic scale of 
a Gaussian distribution by inverting 
Equation~\ref{genusamp} to calculate $\langle k^2 \rangle^{1/2}$
from the amplitude $G_{\rm s,peak}$ of the genus curve.

Even if the distribution is not exactly Gaussian random-phase, 
the maximum amplitude of the genus curve can still give information 
about its characteristic scale at higher densities 
so long as the distribution remains approximately Gaussian
around those higher densities. 
As an analogy to illustrate this argument, 
consider a sea sponge immersed in water. 
When the sponge is moved to dry land, the topology of the enclosed volume, 
evaluated at the density of water, 
will change drastically as water drains and evaporates away. 
At the density of the sponge itself however, 
there will be little change in the topology compared with 
when it was immersed in the sea. 
Similarly, even as ionized bubbles begin growing in the simulation box,  
we can still use the amplitude of the higher 
density regions as an indication of scale 
so long as the bubbles do not occupy a major fraction of the volume. 

\begin{figure}
\plotone{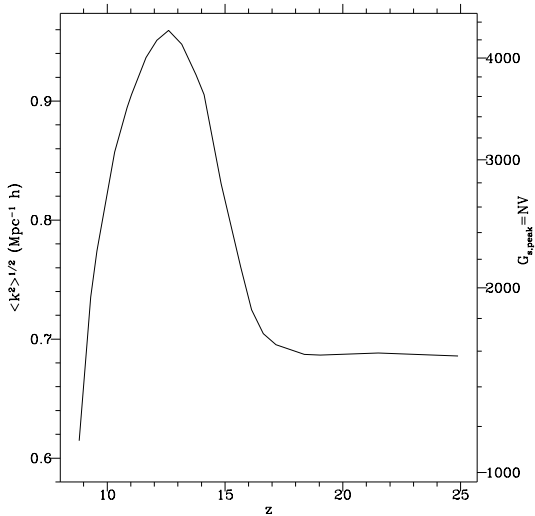}
\caption{The r.m.s wavenumber of the HI distribution, 
plotted as a function of redshift. 
The right axis labels f the corresponding
genus amplitudes $G_{\rm s,peak}$ for our simulation 
volume (see Equation \ref{genusamp}).   
 We truncate the plot at around the epoch when the 
topology departs strongly from a Gaussian distribution,
 at $z \sim 8$ (see discussion in text).}
\label{ampplot}
\end{figure}

A cursory glance at Figure \ref{genuscurves} shows that 
even when the topology deviates from the W-shaped Gaussian curve,
if we ignore the negative $\nu$ (i.e. low-density) part of the genus curves, 
the high-density part does indeed 
retain an approximately Gaussian shape down to $z \sim 10$. 
For example, the $\nu \gtrsim -1$ portion of the $z= 12.1$
curve looks like part of an off-center Gaussian curve.
Immediately after $z\sim 10$, 
the topology of the distribution becomes completely non-Gaussian 
(e.g.\ $z= 8.2$ curve in Figure~\ref{genuscurves}) and 
it becomes difficult to define the amplitude of the peak.
 We plot the genus amplitude $G_{\rm s, peak} = NV$
(for a box volume $V=100^3\; {\rm Mpc}^3 h^{-3}$) and 
the corresponding $\langle k^2 \rangle^{1/2}$ 
at redshifts $9 \lesssim z \lesssim 25$ in Figure \ref{ampplot}.

\begin{figure*}
\plotone{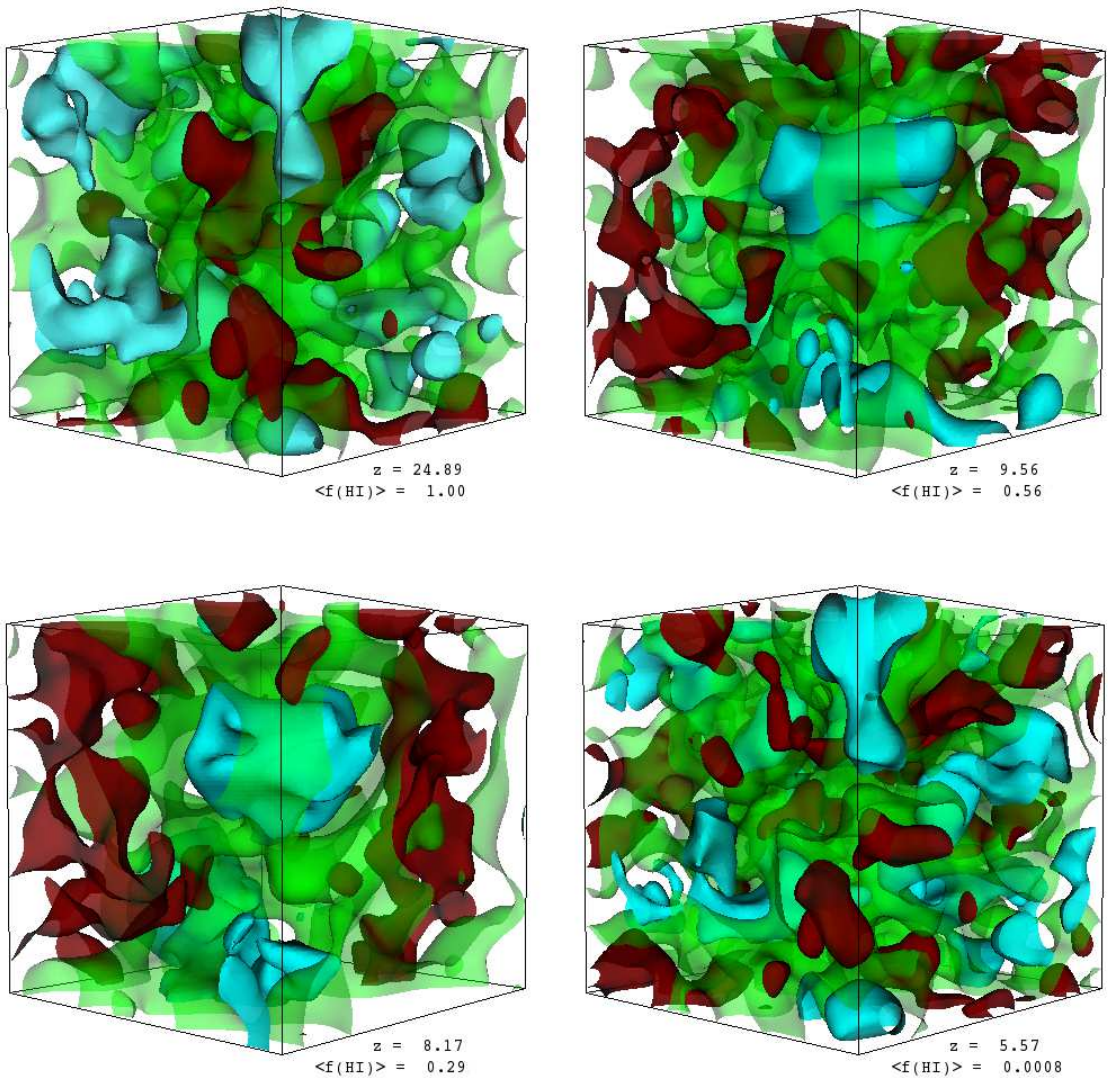}
\caption{HI density contours from a $25 \mpc$ subsection of the overall $100 \mpc$ simulation box, 
labeled by redshifts and their corresponding 
volume-weighted mean neutral fractions. The different colors correspond to densities occupying different volume fractions at that redshift. Blue: $f=0.93$ or $\nu=-1.5$; green: $f=0.5$ or $\nu=0$; red: $f=0.07$ or $\nu=1.5$. Note that actual threshold density values vary with redshift (see discussion in text).}
\label{contourplots}
\end{figure*}

\section{Discussion}
In this section, we present a detailed analysis of the different phases of reionization identified from the variation in the topology of the HI density. In particular, we have identified processes corresponding to `pre-overlap', `overlap' and `post-overlap' reionization first described by \citet{gnedin00a}.  

\subsection{Pre-reionization; $z \gtrsim 16$; $ \bar{f}_{\rm HI} \gtrsim 0.96$}

As the pre-ionization universe is completely neutral at $z \sim 25$, the topology of the HI distribution is that of the primordial Gaussian random phase distribution (top-left panel of Figure \ref{genuscurves}). 
The HI density contours of this distribution are 
illustrated in the top-left panel of Figure \ref{contourplots} 
(for clarity, we only show a $25 \mpc$ section extracted 
from the overall $100 \mpc$ box). 
Both high- and low-density contours are split into 
similar numbers of isolated regions corresponding to a negative genus, 
while the median density contour at $\nu=0$ or $f=0.5$ is 
essentially one multiply-connected object with 
many holes in it, giving a positive genus.   

 While star-formation is already occurring during this epoch, the star-formation rate $ \mathrm{SFR} \lesssim 10^{-3}\, {\rm M}_{\odot}\, \mathrm{Mpc}^{-3} \mathrm{yr}^{-1}$ is too small to provide significant ionizing flux to the IGM; 
we see from Figure~\ref{fHIplot} that 
$\bar{f}_{HI}\approx 1$. 
Hence, the dominant effect on the neutral gas distribution in this era is linear gravitational evolution. 

In Figure \ref{rhoc}, we see that throughout this epoch 
the HI threshold density at $\nu=0$ or $f=0.5$
remains almost the same as the mean density of all hydrogen in the universe,  $\rhoth  \approx 1$. 
Indeed, as the universe evolves during this time, the densities of regions with $\nu > 1$ actually increase relative to the early conditions at $z \sim 25$ while the densities of the voids with $\nu < 1$ decrease slightly. 
Both curves however remain straight lines. 
This behaviour is as expected from linear evolution: the dense regions cluster due to gravitational interactions at the expense of the underdense voids. 
This leaves the topology essentially unchanged, 
as can be ascertained by comparing the plot for $z = 17.2$ 
in Figure~\ref{genuscurves} with that of the 
original conditions at $z= 25$: they are almost identical.

The r.m.s wavenumber of the HI structures also remains approximately constant at $ \langle{k^2}\rangle^{1/2} \approx 0.7\, \mathrm{Mpc}^{-1}\; h$ (corresponding to a characteristic scale  $\langle \lambda^2 \rangle^{1/2} \approx 9 \mpc$) during this epoch as can be seen in Figure \ref{ampplot}. 
Again, this is characteristic of linear evolution, which does not change the topology in any way.

At the end of this phase, we begin to see a rise in the
amplitude of the genus (see plot for $z = 15.7$ in 
Figure~\ref{genuscurves}). 
This is an indication that the reionization is becoming more significant
as HII bubbles begin to grow, leading to the next phase.

\subsection{Pre-overlap; $ 10 \lesssim z \lesssim 16$; $0.63 \lesssim \bar{f}_{\rm HI} \lesssim  0.96$ }

Starting at a redshift of $z \sim 15$, the topology 
starts to change appreciably as reionization begins in earnest. 

The genus curves (e.g. the plot for $z = 15.7$ and
  $z= 12.1$ in Figure \ref{genuscurves})
 start to display an asymmetry between low- and high-$\nu$ portions: 
while the absolute value of the genus for intermediate 
and high density regions ($\nu > 0$)
increase along with the overall amplitude, 
the curve flattens out at low $\nu$, and 
the absolute genus values at these densities decrease. 

A genus curve with this shape is not new: 
when \citet{gott87} studied the topologies of 
non-random phase distributions for the universe, 
one possibility they considered was one in which 
galaxies are distributed at surfaces surrounding large voids.
This has striking similarities with the pre-overlap
genus curves in the present case, 
indicating clearly that large ionized HII bubbles 
are being created during this epoch. 
However, in the present work, 
the bubbles have continuously varying HI densities 
in contrast to the uniform voids studied in \citet{gott87}. 
Indeed, the increase of the absolute value of the genus towards 
the median ($\nu=0$) in the $\nu < 0$ regime is indicative of 
the intuitive fact that the most highly ionized bubbles 
are rarer than less ionized ones, 
thus less bubbles are detected at the lowest threshold densities. 
Also, the topology at high densities is different in both cases. 
In \citet{gott87}, 
the galaxies were distributed randomly on the bubble walls 
(and hence approximately uniformly), 
while in the present case of pre-overlap reionization 
the high density regions outside of the ionized bubbles 
still have an approximately Gaussian distribution of neutral hydrogen. 

From Figure \ref{rhoc}a, we see that the curves relating
threshold density $\rhoth$ to a given volume fraction $f$ or $\nu$
 have a different shape during this epoch, 
with a straight line at $\nu \gtrsim 0$, but with a tail dropping off steeply towards low HI densities at low $\nu$, 
a characteristic which becomes increasingly marked with decreasing redshift. 
This can also be seen in Figure \ref{rhoc}b, 
in which the threshold density for low volume fractions 
begin decreasing rapidly at around $z\sim 14$, 
while the threshold densities at median and 
high volume fractions decrease more gently. 
This is indicative of the drastic reduction in 
neutral density within the interiors of HII bubbles.


In Figure \ref{contourplots}, a comparison between the contours 
at $z = 24.9$ and $z= 9.6$ reveals that regions
initially occupied by high density contours have now become 
low density voids. 
For example, it is clear that the large high-density pocket near 
the middle front of the box in the initial conditions 
has become a low-density void. 
This is a strong indication of `inside-out' reionization. 
In this picture, the highest density regions are the first in which halos can 
accrete sufficient mass (and hence virial temperatures of $T \gtrsim 10^4$) 
for atomic cooling to occur and cause the halos to collapse and commence 
star formation. 
These star-forming regions in turn becomes sources of 
ionizing flux, which ionize the surrounding neutral gas. 

 This effect means that the HII bubbles detected in the 
$\nu < 0$ part of the genus curve are in high-density regions
which have been strongly ionized, 
and not regions with intrinsically low density.

During this stage, the ionizing fluxes are shielded from the IGM by virtue
of being within dense regions with high recombination rates,
 hence the universe outside of the 
HII regions is little affected by reionization.
Outside of the increasingly low HI densities in the HII bubbles, 
the relative densities are still the same. 
For example, a $\nu=-1.5$ void in the $z=24.9$
distribution is by definition less dense than a $\nu=0$ median
density region during the same era, but
they still have approximately the same relative
densities with respect to each other
during the pre-overlap stage of reionization
as they experience little ionizing flux. 
The main difference is that the original 
dense pockets have become HII regions and are now voids
with respect to the rest of the universe, 
so the volume fractions covered by the un-ionized regions
have been reduced (i.e. $\nu$ increased). 
This can be seen by looking at the positions of the 
intermediate (green, $\nu=0$) and dense (red, $\nu=1.5$)
density contours for $z=9.6$ in Figure~\ref{contourplots}:
the former are in regions formerly occupied by the voids 
while the latter are in regions once part of the 
sponge-like median density contour. 
 The correspondence between the contours in is not exact because 
the actual threshold density values for e.g.\ the $\nu=1.5$
contour at $z=9.6$ and $\nu=0$ at $z=24.9$ are not
exactly matched to each other, 
but is sufficient to demonstrate the broad trends
at different densities.

This argument is supported by the evolution of the threshold densities 
with redshift in Figure \ref{rhoc}a. We see, in this context, 
that the low-density tail signifying HII bubbles 
must be from regions that used to be in 
the high-$\nu$ part of the curve in earlier times. 
The straight portion at $\nu \gtrsim 0$ is from regions 
that once occupied the low density $\nu \lesssim 0$ portion 
of the straight line representing the Gaussian distribution 
before reionization began. These have now been translated to occupy higher 
volume fractions relative to the overall distribution. 
However, even in these regions, 
some small amount of ionization is taking place 
at scales below the smoothing length, 
which slowly decrease the local HI densities. 

As HII regions begin forming in the regions of 
highest baryon density, recombination initially mitigates
the effects of the ionizing flux. 
Thus, these partially ionized regions initially appear 
as bubbles of intermediate neutral density 
growing within highly dense regions. 
This has the effect of creating holes and 
isolating individual dense clumps 
within the dense regions as the HII regions appear, 
breaking up the topology and increasing the amplitude of the genus.

As reionization progresses, the absolute genus value of 
the low-density regions decreases due to the merger of 
small HII bubbles into a smaller number of large bubbles 
(compare $\nu < -1$ tails of $z=12.10$ and 
$z=9.6$ in Figure~\ref{genuscurves}).
This is in agreement with \citet{shin07}, 
which found that the majority of bubble mergers occur during this epoch.

\subsection{Overlap; $ 8 \lesssim z \lesssim 10$; $0.26 \lesssim \bar{f}_{\rm HI} \lesssim 0.63$}

At $ z \sim 9$, volume-weighted mean HI fraction of the universe 
 drops to $\bar{f}_{\rm HI} \lesssim 0.5$ (Figure \ref{fHIplot}), 
in contrast with the leisurely decrease in previous phases,
 even though there is merely a modest increase in the amount of ionizing flux 
emitted by the star-forming regions 
(now dominated by Pop II stars; see \citet{trac06b}). 
This sudden decrease occurs as the ionized bubbles overlap with each other, 
so that ionizing photons can now stream between previously isolated bubbles 
and increase the ionizing flux experienced 
by points within the overlapping ionized regions.
This increased flux overcomes the effects of recombination 
and increases the ionization rate. 
At regions occupying the lowest volume fractions, 
the gas is essentially completely ionized, 
with $\rhoth \lesssim 0.01$ at $\nu= -1.5$ (Figure \ref{rhoc}b). 

As the regions occupying the median 
volume fractions ($f=0.5$ or $\nu \sim 0$) are 
now also part of significantly ionized regions, 
the genus of what were previously multiply-connected regions is 
now strongly suppressed due to the overlapping of HII regions, 
which closes gaps in the sponge-like contours 
(and hence reduces the positive genus).
By $z= 9.6$, the overall amplitude
of the genus curve is considerably lower than
that at $z= 12.1$ (Figure~\ref{genuscurves}). 
This reflects an increase in the characteristic scale 
of the neutral distribution,
driving down the r.m.s. wavenumber and 
hence the genus amplitude (Figure~\ref{ampplot}). 
Immediately after $z \sim 9$, 
the topology of the distribution across all densities becomes
strongly non-Gaussian as can be seen from a comparison
of the genus curves for $z = 9.6$ 
and $z = 8.2$ in Figure~\ref{genuscurves}. 
After this point, there no longer exists large neutral structures
with approximately Gaussian distributions within them.
This means we can no longer estimate a characteristic scale 
using Equation~\ref{genusamp} and so we truncate Figure~\ref{ampplot}
at $z\sim 9$.

The overlap of ionized regions means that the topology 
at low-densities is dominated by a small number 
of large interconnected regions, 
which is detected as $G_s \gtrsim 0$ in the $\nu<0$ portion 
of the genus curve for 
$z = 8.2$ (Figure \ref{genuscurves}),
 and is seen in the corresponding contours in 
Figure \ref{contourplots}.

With the percolation of HII bubbles through the IGM, 
the remaining regions with a relatively high HI density 
($\rhoth \gtrsim 0.5$ at $\nu > 0$) are now isolated into 
individual pockets (see the corresponding contours in Figure \ref{contourplots}). 
This is also seen in the high-density $\nu \gtrsim 1.5$ portion 
in the genus curve for $z= 8.2$ in Figure \ref{genuscurves}, 
which retains vestiges of its curved Gaussian shape. 
This is because these high density regions are still shielded 
from the ionizing flux flooding the rest of the universe, 
thus within them the {\it relative} density distributions 
still reflect those of the initial conditions. 
But even within these pockets, small amounts of star formation are going on 
at scales below the smoothing length, 
which continue pushing down slowly on the densities. 
The stage is now set for the reionization of the remaining large-scale HI structures in the universe.  

\subsection{Post-overlap; $6 \lesssim z \lesssim 8$; $ \bar{f}_{\rm HI} \lesssim 0.26$}

By redshifts of $z \sim 8$, the median HI density of universe is $\rhoth \sim 0.1$ and dropping (Figure \ref{rhoc}). 
The contours with threshold densities below the median, 
$\nu < 0$, are now almost thoroughly ionized, as characterised by the flat 
tails in the corresponding $\rhoth$ v.s. $\nu$ plots in Figure \ref{rhoc}a.  

In the genus curve for $z = 7.3$ (Figure \ref{genuscurves}), 
the genus at $\nu < 0$ has now assumed a shape 
that appears similar to that of a Gaussian distribution, 
but truncated where it crosses the median volume fraction at $\nu \approx 0$. 
This indicates that regions below these 
threshold densities are totally ionized, 
with the unresolved neutral regions (i.e. galaxies) within them 
acting as tracers of the baryon's Gaussian random phase density distribution. 
A comparison with the curve in Figure \ref{rhoc} at this redshift concurs 
with the notion that a straight line in the $\rhoth$ v.s. $\nu$ curve 
is indicative of a random phase distribution. 
At $0 \lesssim \nu \lesssim 1.5$, ionized regions have yet to be completely
 ionized, and have the topology of a unified region punctuated 
by small numbers of holes. 

These holes are caused by the remaining resolved pockets of high-density 
HI in the universe, which by this epoch has a density 
about 2 orders of magnitude greater than fully ionized regions 
occupying most of the IGM (Figure~\ref{rhoc}b). 
These pockets can also be seen in the genus curve at $\nu \gtrsim 1.5$.
 However, the small negative amplitude of the genus 
indicates that only a small number of neutral pockets remain, 
and corresponding volume fraction 
they occupy is small ($f = 0.02$ at $\nu = 2$).  

With the universe now largely ionized, 
most of the IGM has a low optical depth to ionizing photons. 
The remaining HI regions are isolated pockets occupying small volumes, 
which implies that they have relatively large surface areas 
exposed to large amounts of ionizing flux.
Under these circumstances, the final reionization of 
these pockets happen within a relatively short period of time 
as illustrated in Figure \ref{rhoc}a, 
where $\rhoth$ at  $\nu \sim 1.5$ decreases by about 2 orders of magnitude 
during the interval $5.6 \lesssim z \lesssim 7$. 

The genus curve at $z = 6.4$ (Figure~\ref{genuscurves}) 
offers an interesting snapshot of the topology during this time. 
The curve at $\nu < 1$ is distinctively that of the 
Gaussian random-phase distribution, corresponding to the 
baryonic distribution traced by the unresolved neutral HI. 
At $\nu > 1$, we see that the densest $\sim 10 \%$ 
($f \sim 0.1$) of the volume are in a very tiny number
of dense neutral pockets, but they no longer
exist by $z = 5.6$.

At the end of post-overlap reionization at $z = 5.6$, 
the HI threshold density is a straight line 
with respect to $\nu$ (Figure \ref{rhoc}a), 
with $\rhoth \lesssim 0.01$. 
This indicates that the reionization of the universe is complete 
by this redshift, and the tiny amounts of neutral HI remaining in 
the universe now traces the overall baryon distribution in the universe, which 
is still Gaussian random phase to a good approximation. 
This is supported by the corresponding genus curves (Figure \ref{genuscurves})
and contours (Figure \ref{contourplots})).

We note that a post-reionization 
Gaussian random-phase distribution
 is directly supported 
by observational evidence: \citet{weinberg98} had studied 
the 1-dimensional topology of neutral hydrogen
 using Ly-$\alpha$ absorption lines in quasars 
with $z \lesssim 2.5$, and it was found that 
the distribution of threshold crossings 
as a function of $\nu$ (i.e. number of threshold crossings $\propto \exp{-\nu^2/2}$) smoothed on scales of $1 \mpc$ is consistent with a Gaussian random-phase distribution .

In the contours plots, the distribution of 
the high- and low-density regions at 
$z = 5.6$ bear a resemblance to the 
initial conditions at $z = 24.9$. 
For example, the anvil-shaped void at the top vertex facing the reader can 
clearly be matched with an similarly-shaped object in the initial conditions,
 as is the small void at the lower vertex. 
In general, the contour distributions in the initial and post-reionization 
volume can be approximately matched to each other, but there are small differences. 

The post-reionization universe is in general choppier than the 
initial conditions, most noticeably in the dense regions. 
This is supported by a comparison between the two corresponding genus curves 
(Figure \ref{genuscurves}). 
The amplitude of the genus curve at $z = 5.6$ is about twice that at $z = 24.9$, giving a corresponding decrease in the characteristic wavelength according to Equation \ref{genusamp}. 
This difference is most likely due to non-linear effects
which occurred in the baryon distribution whilst
reionization was taking place, 
but is beyond the scope of this paper.

\section{Conclusion}
We use a state-of-the-art cosmological 
reionization simulation to quantify the topology of cosmological reionization,
using genus curve measurements of the neutral hydrogen density distribution on $\ge 1 \mpc$  scales.
Overall, the reionization process begins
in an inside-out fashion on scales,
where higher density regions become ionized earlier than lower
density regions. 
Reionization finally completes when 
the isolated neutral structures which remain are
reionized in from outside-in.

The sequence of topological transitions can be mapped to four phases: 

(1) pre-reionization: $z \gtrsim 15$, 
(2) pre-overlap: $10 \lesssim z \lesssim 15$,
(3) overlap: $8 \lesssim z \lesssim 10$,
(4) post-overlap: $6 \lesssim z \lesssim 8$.

In Phase (1) the genus curve is consistent with a Gaussian
density distribution and topology remains unchanged as the universe evolves via linear evolution.
In Phase (2) the number of HII bubbles increases gradually with time,
coincident with the increase in the amplitude of the genus curve. Most of this occurs in the highest density regions which begin star-formation earlier.
At the later stages of this phase,
 HII bubbles start to merge/percolate,
resulting in a turn-over and subsequent decrease in the amplitude 
of the genus curve.
An isolated HII-bubble dominated topology is evident 
in the genus curve during this phase.
In Phase (3) the HII bubbles rapidly merge,
manifested by the precipitous drop in the amplitude of the genus curve. 
The reionization of the IGM occurs rapidly as the ionizing photons 
are now free to stream freely between connected HII regions, 
increasing local ionizing fluxes.
In Phase (4) the IGM is thoroughly ionized
and the genus curve is consistent with an increasingly diminishing number
of isolated neutral islands, 
with the reionization of the universe at $\sim 1 \mpc$ 
complete by a redshift of $z \approx 6$.

These results represent a concrete identification of the 
redshifts at which the different epochs of reionization occur, 
and elucidates the evolution of the neutral topology.

\acknowledgments
This work is supported in part by grants AST-0407176 and NNG06GI09G. HT is supported in part by NASA grant LTSA-03-000-0090. JRG is supported by grant AST-04-06713. We thank Min-Su Shin for useful discussions and advice. 
We are grateful to the referee, Nick Gnedin, for his constructive criticisms and suggestions which have improved the paper.



\end{document}